# A FRAMEWORK FOR ASPECTUAL REQUIREMENTS VALIDATION: AN EXPERIMENTAL STUDY

Abdelsalam M. Maatuk[1], Sohil F. Alshareef [2] and Tawfig M. Abdelaziz[3]

[1&2]Department of Software Engineering, University of Benghazi, Benghazi, Libya
[3]Libyan International Medial University, Benghazi, Libya

## *ABSTRACT*

*Requirements engineering is a discipline of software engineering that is concerned with the identification and handling of user and system requirements. Aspect-Oriented Requirements Engineering (AORE) extends the existing requirements engineering approaches to cope with the issue of tangling and scattering resulted from crosscutting concerns. Crosscutting concerns are considered as potential aspects and can lead to the phenomena "tyranny of the dominant decomposition". Requirements-level aspects are responsible for producing scattered and tangled descriptions of requirements in the requirements document. Validation of requirements artefacts is an essential task in software development. This task ensures that requirements are correct and valid in terms of completeness and consistency, hence, reducing the development cost, maintenance and establish an approximately correct estimate of effort and completion time of the project. In this paper, we present a validation framework to validate the aspectual requirements and the crosscutting relationship of concerns that are resulted from the requirements engineering phase. The proposed framework comprises a high-level and low-level validation to implement on software requirements specification (SRS). The high-level validation validates the concerns with stakeholders, whereas the low-level validation validates the aspectual requirement by requirements engineers and analysts using a checklist. The approach has been evaluated using an experimental study on two AORE approaches. The approaches are viewpoint-based called AORE with ArCaDe and lexical analysis based on Theme/Doc approach. The results obtained from the study demonstrate that the proposed framework is an effective validation model for AORE artefacts.*

## 1. INTRODUCTION

The Aspect-Oriented Requirements Engineering (AORE) aims to extract and construct good identification and separation of crosscutting concerns [14]. Conventional requirements engineering (RE) approaches treat the resulting intersection of concerns, i.e., aspects, as a rule, or restriction relationship [33] [6]. Such concerns handling might lead to what so-called tyranny of the dominant decomposition, which results in scattered and tangled implementation of concerns to other software artefacts [16]. In terms of aspect orientation [6][16], traditional RE approaches have been extended to support the crosscutting nature of concerns as early as possible in software development.

In this context, shortcomings, such as clean mapping of requirements artefacts to later stages, satisfactorily trade-off analysis, equality treating of concerns and validation of AORE artefacts remain a challenge and not fully studied [1][13].

Validation in requirements engineering is the process to eliminate the conflicts among requirements in SRS [1][2][41]. Besides, it relates to the analysis since it is involved with detecting problems associated with the SRS, i.e., inconsistencies, incompleteness and ambiguities [2] [43]. Inconsistent requirements are requirements that are stated and described differently at different places in SRS, whereas ambiguous requirements referred to as one requirement has multiple perceptions and incomplete requirement, which refers to a requirement that does not convey a complete and meaningful requirement [43][44].

Furthermore, requirements validation is a crucial activity as it helps to reduce the cost of maintenance. This is due to the reason that, the cost of fixing errors in later stages, i.e., architecture, design and code, of software development is more expensive than rectifying them in the requirements analysis phase [2]. In addition, errors in SRS lead to huge rework costs and effort when issues are discovered in later stages. Fixing errors in later stages of software development requires more time, effort and cost [33][15][42]. Moreover, making changes at this stage requires requirements engineers and analysts to properly manage the changes as they would affect other artefacts, which are associated with the requirements and it would to re-testing of the system when the new changes are employed.

Through the investigation described in [3], it is noticed that almost all the approaches focus on identifying crosscutting concerns as well as representing them either using graphical models, XML schema, templates or text-based representation. Therefore, the resulting artefacts from the requirements phase must be validated before moving to other development stages to reduce extra work and effort caused by missing information and unintentionally overlooked inconsistency rules [43]. In addition, the validation can be either dynamic or static. Dynamic validation is easier to perform and it allows the automatic checking and interpretation of software artefacts, e.g., Test Driven Development. In contrast, static validation implies the understandability and completeness of requirements artefacts to be validated. However, dynamic validation is often more costly than static validation as the problem domain needs to be formalized and fully covered at first.

As far as we can tell, the AORE proposals lack the clarity to employ techniques and methods to validate aspectual requirements after they are identified to ensure the correctness of resulted AORE artefacts. Some experiments and case studies were conducted in terms of aspects and concerns identification. Therefore, we have proposed a framework, i.e., Validation Framework for Aspectual Requirements (ValFAR) to validates the AORE artefacts in three main phases, each of which has several activities [12]. Phase 1 is for concern handling, Phase 2 is for concern validation, i.e., high-level validation, whereas Phase 3 consists of Aspectual requirement specification and validation, i.e., low-level validation. However, the focus of this paper is mainly on the experimental work of the framework and the validation problem of aspects and concerns in the requirements engineering phase. The results of the experimental work are presented and discussed. The data used in the experiments were the resulted analysis artefacts from two different approaches.

The correctness of the ValFAR method has been tested with analysis artefacts from existing AORE approaches. These artefacts are used as input to ValFAR to ensure that the resulted concerns and aspects are valid and correct. Two different AORE approaches have been selected to apply the proposed solution. On one hand, the first approach is a viewpoint-based model [6], which treats aspectual requirements as concerns and encapsulates them in a simple self-explanatory XML schema. The schema, however, contains requirements, e.g., concerns, which should be decomposed to sub-concerns to reduce the tangling issue and make it simpler to apply the decomposition rules. On the other hand, the second approach is a lexical-based aspect identification, i.e., Theme/Doc [5][10]. It uses views and action-view to identify the relationships between concerns and requirements to extract the influential aspect. The Theme is suitable when the requirements are well-written and documented, although it cannot identify non-functional concerns as they are usually not associated with action words. The results from both experiments were encouraging, although some artefacts had to be refined and modified in terms of rewriting descriptions of concerns and requirements to validate AORE artefacts.

The remainder of the paper is organized as follows: Section 2 presents the related work to the validation and verification of aspects and crosscutting concerns. Section 3 introduces the ValFAR method. Section 4 presents the experimental study. Section 5 discusses some of the obtained results. Section 6 concludes the paper.

## 2. RELATED WORK

Different AORE approaches are dedicated to aspectual requirements focusing on different stages during the system development. However, through our investigation of the problem, we have found that only a few AORE proposals provide the means to validate the elicited aspects [6][11][7]. A comprehensive survey for aspect mining tools and approaches can be found in [40]. The work presented in [36] describes the synthesis and interaction of aspects in aspect-oriented methodology.

The theme approach [5], is based on lexical analysis of the requirements and it has proven to be an effective method to identify and extract the relationship among aspects by action verbs. AORE with ArCaDe [6] identifies aspects as viewpoints and it encapsulates them in an XML schema with their requirements and sup-requirements. However, this method focuses on the composition of aspectual requirements rather than validating the concerns and the identified aspects.

The AORE model proposed in [6] applies the viewpoint-oriented approach to identify stakeholders' requirements using viewpoints and XML. Araŭjo et al. [6] introduced a framework based on a viewpoint-oriented approach, which generates templates to identify crosscutting non-functional requirements.

Grundy [8] proposed an aspect-oriented requirement engineering solution for component-based systems, whereas Jackson [9][13] proposed a problem frame approach, which identifies aspects from the problem domain. This method is only applicable when the problem being investigated has been solved in the past, and the solution can be used again, by breaking down problems into sub-problems in parallel with defining composition rules, which can be used to resolve the conflict between aspects.

Other models, e.g., Cosmos [13] and CORE [7] are proposed for concern-oriented or multidimensional separation of concern approach. Although Cosmos is known as the best for its ability to analyze the relationship between concerns, it lacks the systematic means to identify any type of concern. Moreira [7] proposes an approach, called CORE that introduces a projection table to reduce the concern dimension.

The Theme approach [5][10] is an AORE approach that is not based on conventional requirements engineering techniques, such as viewpoint, use case scenario and goal-oriented. Theme solution extracts crosscutting concerns and identifies aspects by keyword and lexical analysis.

Araŭjo et al. [16] proposed a use-case-based requirement approach, which identifies concerns from different scenarios of the system. It is a template-based approach for identifying aspectual use-cases, which are modelled, providing traceability similar to the multidimensional separation of concerns techniques. In addition, there are many approaches to identify aspects by use-case [14][15][16].

Domain analysis of existing component and aspect approaches for crosscutting concerns is described in [39]. They identify a set of requirements to be fulfilled for modeling of dynamic and crosscutting features using aspect-oriented. After that, the identified requirements are grouped into an evaluation model to be checked against capability of existing combinations to validate the model.

A process that identifies the relations that can be observed between architectural aspects, quality attributes, and requirements is presented in [38]. The relations are created with the aid of mappings from non-functional requirements to quality attributes and architectural aspects. The process uses the MultiCoS approach in managing requirements and concerns by matching function and crosscutting coefficients to identify the crosscutting concerns.

In terms of identification, an aspect frame structure is proposed, which is a pattern for aspects that share a common concern, behavior, and how are integrated into a rational of the functional requirements, which they crosscut [37]. These frames are used to describe the concrete aspects by requirements engineers and analysts.

In the context of aspectual requirements validation, we have found only one framework presented in [4], which argues that the validation of aspectual requirements can be accomplished with two levels of validation. It amalgamates three AORE approaches: viewpoint-oriented approach, goal-oriented approach and a use case-oriented approach. It converts the identified concerns from each AORE approach into their respective graphs. An aspectual graph is produced and provided as an input to the aspectual graph parser, which generates an XML file for the aspectual graph to compare it with each XML file from each approach to remove any duplicated concerns and ambiguities. However, there is no way to validate the framework in general. It was presented without actual results or case studies to prove that the method is working properly with these three approaches. The conversion process itself is not clearly defined with specific steps to follow neither the integration of the three models into one XML file.

Validation of requirements artefacts in traditional requirements engineering is supported by using one of the common validation techniques, e.g., formal review, inspection, or paper prototyping. In addition, there is a separate activity involving validating the requirements in traditional requirements engineering. In contrast to that, AORE approaches are focused more on the identification and specification of aspects and concerns without equally giving attention to the test and validate those artefacts. However, the analysis techniques presented by each model provide a great granularity of information to conduct the validation. The challenge was the ability to use one validation method on different approaches without the need to change the validation to cope with each AORE model. This paper introduces a novel method to validate and evaluate aspectual requirements in the early stages using experimental study.

## 3. THE PROPOSED APPROACH

This section describes the proposed solution, i.e., the Validation Framework for Aspectual Requirements (ValFAR) [12]. It validates the AORE artefacts in three main phases, each of which has its activities as shown in Figure 1. Phase 1 is for concern handling, Phase 2 is for concern validation, i.e., high-level validation, whereas Phase 3: consists of Aspectual requirement specification and validation, i.e., low-level validation. These phases will be described in the subsequent subsections. Before Phase 1, the correlation between concerns that form the crosscutting relationship should be structurally addressed with a matrix. In addition, it is useful to link the source of information of each concern with the corresponding stakeholder.

More importantly, ValFAR does not identify concerns or crosscutting concerns. It operates on the assumption that AORE artefacts are identified in the form of the used approach, e.g., goal-oriented (soft-goal and sub-goals), viewpoint-oriented, themes. However, before using the ValFAR, these conditions have to be met:

- The SRS, i.e., the requirements document is produced
- Crosscutting concerns are identified using an AORE approach
- Aspectual requirements and concerns are specified in the SRS

## 3.1. Phase 1: Concern Handling

AORE approaches treat concerns based on the approach being used to identify concerns and aspects. In a viewpoint-oriented approach, concerns are treated as viewpoints. Goal-oriented approaches are based on soft-goal and sub-goal concepts and aspects are discovered and identified as operationalities that affect the sub-goal, i.e., non-functional concern. In the Theme/Doc approach, the theme and base theme is functional and non-functional concerns respectively. Thus, we propose this phase to treat each artefact, i.e., viewpoint, goal, sub-goal, base theme and crosscutting theme, which is used to identify aspects, like a concern to perform the activities to achieve high granularity. This phase comprises three activities described as follows:

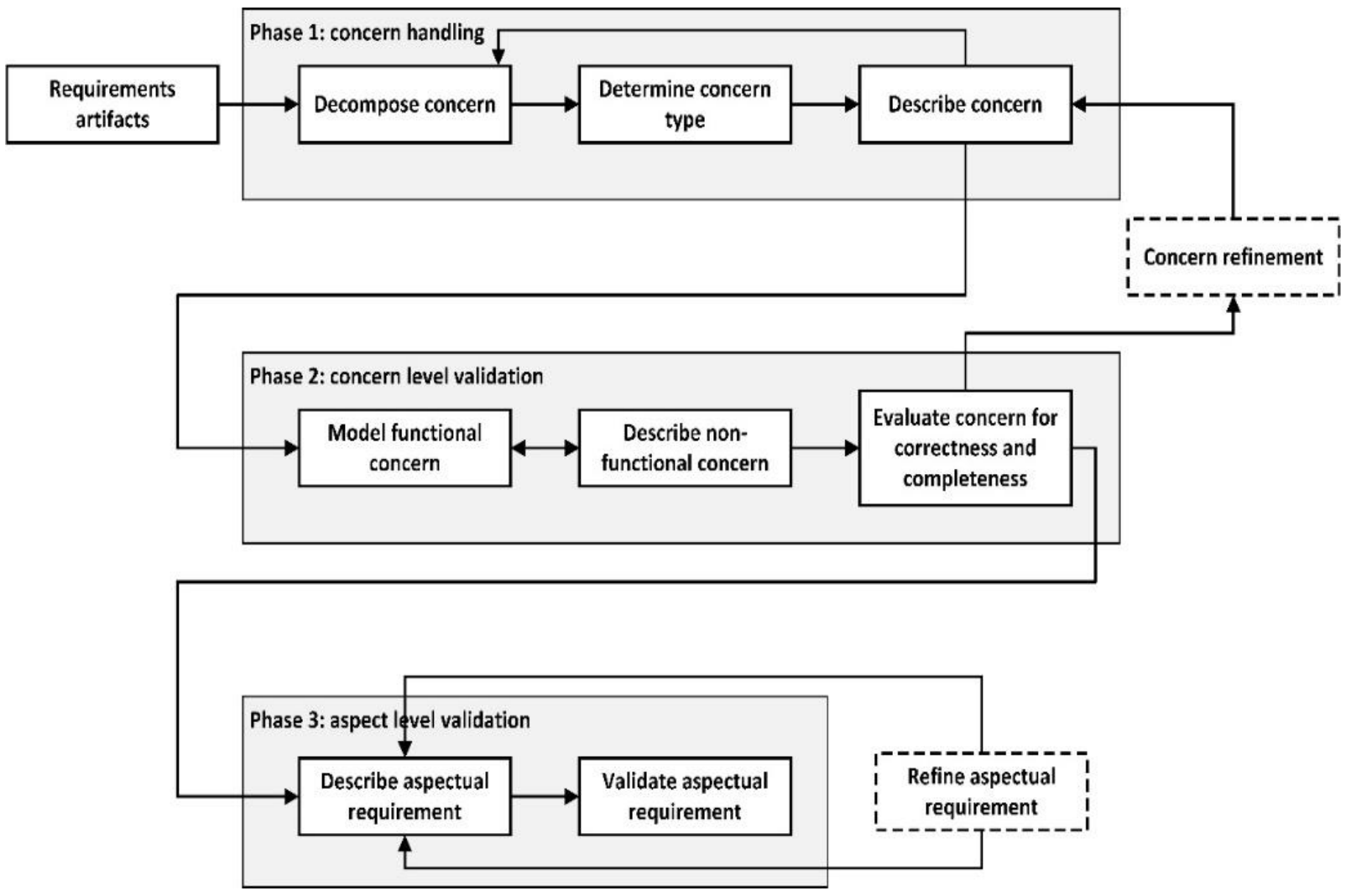

Figure 1. illustrates the process of the ValFAR

### 3.1.1. Activity 1: Determine Concern Type

The Concern type should be determined before processing. A concern might be functional or non-functional. Reviewing the analysis documents and the requirements specification helps to understand the nature of concerns, hence determining its type. Functional concerns are treated differently from non-functional in terms of specification, modelling. Thus, determining concern type is crucial for granularity.

#### 3.1.2. Activity 2: Determine Concern Type

Concerns often have tangled nature in their descriptions, which poses as a difficulty to analysts to handle them uniformly [20]. In addition, these entanglements contain additional information that should be separated or new ones that are derived but encapsulated when needed [19][21][22]. To tackle this difficulty, and if one or more of the rules are met, concerns should be decomposed based on the following rules:

- If some parts of the concern present a limitation as to how a specific goal is achieved
- If there are parts implied design constraints on the original concern.
- If there are parts of the concern that are not used by other concerns, but they are used to describe the issue
- If the description of the issue is too long
- If the concern contains a reference of description to other different concerns
- if a concern requires information from different concerns
- If there is a need to derive new concerns from the original concern
- If there are some parts of the original concern that are going to be associated with other concerns

#### 3.1.3. Activity 3: Describe Concern

Concerns often have a tangled nature in their descriptions, which poses a difficulty to analysts to handle them uniformly [20]. In addition, these entanglements contain additional information that should be separated or new ones that are derived but encapsulated when needed [19][21][22]. To tackle this difficulty, and if one or more of the rules are met, concerns should be decomposed based on the following rules:

- Concern ID: the unique identifier of the concern for traceability purposes.
- Concern name: a name that should not be repeated elsewhere to prevent inconsistencies and confusion.
- Objective: a brief description of concerns objective written in a clear language.
- Type: type of concern, whether it is a functional or non-functional concern.
- Successful scenarios: a concise and complete description of the process executed by the concern.
- Alternative scenarios: an alternative to the main scenario when limitations constraint it.
- Revision date: the date that the concern was modified last.
- Review count: indicates the number of iterations performed to process the concern.
- Reference: a reference to the concern analysis process and concern representation.

### 3.2. Phase 2: Concern Validation (high-level validation)

In Phase 1, concerns are decomposed and described structurally based on the type of concern. Moreover, in Phase 2, functional concerns are validated using a semi-formal technique [2] [35]. In addition, non-functional concerns that constraint the functional concerns are linked to the influenced concerns described and separated in a dedicated template. The stakeholder is part of this phase as the developer validates the functional concerns and gets immediate feedback from the stakeholder. The following three activities describe this phase.

#### 3.2.1. Activity 1: Model Functional Concerns

In this activity, the functional concerns are validated in a semi-formal manner [24]. The UML sequence diagram model is used to verify the interactions with corresponding artefacts of the system [27]. Besides, the specification of the concern must be validated to ensure the correctness and it satisfies its objective. Additionally, the stakeholder, i.e., end-user or customer, is involved for better results [25][26]. The number of iterations of this activity depends on the granularity of the feedback between the customer and the analyst.

#### 3.2.2. Activity 2: Describe Non-functional Concerns

In contrast to functional concerns, non-functional concerns usually are not associated with verbs that express action and cannot be modelled by UML models [29][34]. In addition, non-functional concerns employ how the system should respond and react to the user's interaction, i.e., response time or performance [30] [28]. Therefore, we propose a structured template to describe non-functional constraints or design limitations that limit the structure of the functional concern. The elements to be included in the non-functional template are:

- Name: name of the non-functional concern as described by developers.
- ID: a unique identifier to avoid redundancy and increase the ability to trace and concern.
- Related concerns: functional concerns that are constrained by this non-functional concern.
- Specification: clear and concise description of the limitation of this concern on other concerns.
- Revision date: a date on which the concern was last reviewed or altered.
- Review count: number of iterations performed by developers to process the concern.

A functional concern might be constrained by more than one non-functional concern [29][30]. Using this technique to document and specify non-functional concerns benefits in several ways:

- Developers can derive architectural choices from the template and behavior of the functional concern.
- Separate functional and non-functional concern specifications to improve the understandability and readability of the artefacts.
- Separate overlapping quality attributes that might be included without detailed analysis within the functional concern.
- Improve traceability of non-functional concerns, since the template can be altered at any given time during the development in terms of evolution or refinement.

#### 3.2.3. Activity 3: Evaluate Concerns Correctness and Completeness

Concerns that are part of the crosscutting relationship must be investigated to validate the relationship and the integrity of the aspect through inspecting the concerns using a checklist that contains quality questions aimed at evaluating the completeness and consistency, i.e., logical and grammatical consistency [43]. This activity helps to ensure that concerns and requirements have been extracted well and are ready to be moved to the next stage. In addition, all of the questions of the checklist must be answered as satisfactory to each element; otherwise, the process is repeated until all questions are answered with "Yes" except for missing details or anything forgotten, the answer is "No" [11]. The questions to be included in the template are:

- Is the concern defined clearly?
- Are there any missing details or anything forgotten?
- Is there any association with other concerns or requirements?
- Does it include all the necessary information to make an architectural decision and design it?
- Is there enough information to develop a test case for the concern?

- Is the successful scenario of each concern written in clear language?
- Is the description written so that it won't lead to misinterpretation?
- Are there any parts of the concern repeated in other concerns?
- Are there any conflicts with other concerns or requirements?
- Is the concern traceable to its origins and stakeholder?

## 3.3. Phase 3: Aspectual Requirement Specification and Validation

The crosscutting relationship among concerns is validated according to the defined criteria in Phase 2. The area of intersection, i.e., common interest with other concerns or requirements is considered as a candidate aspect. Thus, aspectual requirements should be specified correctly. Although aspects are systematically extracted with the aid of AORE approaches, the lack of capturing all the necessary information in a single unit structurally, e.g., aspectual requirement document remains overlooked. To overcome this issue, we propose a structured template to document the critical details and the required information to validate the aspect statically as Phase 3 of the solution, which contains two main activities.

### 3.3.1. Activity 1: Describe as an Aspectual Requirement

Aspectual requirements should not be too technical and have to be as abstract as possible [33]. The aspectual requirement describes the solution concept of the crosscutting concern, which has to be carefully treated when designing and implementing the aspect in later stages. Besides, the description of the aspectual requirements has to be clear and concise [31]. The details to be included in the aspect document are:

- Aspect ID: a unique identifier for the aspect.
- Aspect name: a name to describe the aspect and it should not be repeated with other aspects.
- Concerns: list of concerns or concerns that are influenced by this aspect.
- Aspect description: a brief and concise description of the aspect of behavior.
- Aspect priority: the degree of importance for the aspect demanded by stakeholders.
- Aspect post-condition: a service that the aspect will provide.
- Reference: Reference to the analysis material, documentation and representation of the aspect for understandability.

After documenting the aspectual requirement, a matrix is developed if there is a dependency between aspects in terms of execution and responsibility [32]. In order to achieve a good result from this matrix, the analysis of the requirements and concerns must have addressed the aspect life cycle. It could be shown as a graphical representation of the aspect or through tables provided by some of the AORE approaches that handle this issue.

### 3.3.2. Activity 2: Validate the Aspectual Requirement

To ensure the quality of the information acquired by the requirements engineers, we propose to develop a list of questions that are targeting the core concepts of the aspects at the requirements level. In other words, the other development stages, i.e., architecture, design and coding should not be included in the description of the aspectual requirements. Therefore, the requirements section of the validation form must be answered to "Yes". This means that all the necessary information for the aspectual requirements is available in the correct format. The description of the aspectual solution should not limit the ability of developers to design the aspect. However, if there are constraints or design limitations that should be mentioned in the description, developers need to consider that to perform trade-off analysis and negotiation with stakeholders. The questions to be included in the template are:

- Does the aspect describe the solution concept for the problem?
- Does the aspect define the process with an accurate description and understandable language?
- Does the aspect contain all the related parts of the crosscutting concerns?
- Have the areas of intersection with other requirements or concerns have been identified and specified within the scope of the aspect?
- Are the requirements and concerns affected by the aspect identified and represented clearly?

## 4. EXPERIMENTAL STUDY

This section demonstrates the experimental work of ValFAR that is conducted on the analysis artefacts resulted from two existing AORE approaches. These approaches have been chosen for this study due to their nature in handling the AORE artefacts. The Theme is uniquely designed for the analysis and design of aspectual requirements. It uses action words and action clipped view to capture and model the crosscutting relationship respectively, which is unlikely in other approaches that use UML as modelling tool. The AORE with ArCaDe, on the other hand, encapsulates the functional and non-functional concerns in XML schema and treats it as an aspect. This encapsulation presents a challenge in determining the degree of interaction among the inter-related concerns. For these reasons, these two approaches have been selected to demonstrate the ValFAR method. For the lack of space, we have chosen to show the final results of the refinement process for requirements and concerns in Table 4 and Table 10. There are several iterations to reach the desired output. Although, changes have been made to the artefacts without harming the relevancy of the crosscutting relationship. Besides, Tables 2, 6 and 12 have reviewed to the corresponding number in the count section of each table. The refinement of the artefacts was necessary to improve the results obtained from the approaches. Some of the requirements and concerns have been re-written and re-analyzed for better understanding and granularity.

### 4.1. Viewpoint and XML Approach

The case study demonstrated by AORE with ArCaDe is a simplified version of the toll collection system [6]. The approach has identified a number of viewpoints *TollGate, Vehicle, ATM, Police, Gizmo* and *DebitingSystem*. In this paper, the *TollGate* viewpoint will be demonstrated and *ResponseTime*, i.e., an aspect which crosscuts multiple viewpoints, is demonstrated as well. Besides, concerns and requirements are treated as a viewpoint in this approach. The study shows that aspects can be composed and identified using viewpoint methodology. The *TollGate* viewpoint contains tangled information, e.g., exiting and entering the toll, type of toll and capture plate number for unauthorized vehicles. For better granularity, this viewpoint should be decomposed to separate the tangled description of the included requirements. Table 1 illustrates the final results of the decomposition process.

Table 1. Decomposition of TollGate concern

| Decomposed Concern | |
|---|---|
| Core and sub-concerns | Description |
| TollGate <<core cocnern>> | A gate with installed cameras and sensors that vehicle passes through. |
| EntryToll | Detects the installed gizmo on the vehicle. |
| SingleToll | Turn the light into the green for authorized vehicles and display the amount of money to be paid. |
| PayToll | Display the amount of money to be paid by authorized vehicles. |
| PlateCapture | It captures the plate numbers of unauthorized vehicles. |
| ExitToll | Checks the entrance of the vehicle through the gate whether it is a valid entrance or not. |
| Note | The PayToll and PlateCapture concerns are added as new sub-concerns after analyzing the initial scenario of the system. Thus, this refinement should provide a clear and concise understanding of these two concerns. However, when validating the concern, these sub-concerns have to be taken into consideration as a whole. |

Once the type of concerns is determined and are decomposed, concerns, as demonstrated in Table 2, are described in the template as shown in Activity 3 of Phase 1 of ValFAR. This task is only applied to functional concerns to be validated in Phase 2. The functional concerns are modeled and validated using a sequence diagram with the involvement of the concerned stakeholders, i.e., analysts and users. The validation of Vehicle and Gizmo concerns are shown in Figures 1 and 2 respectively

.

Table 2. Description of TollGate concern

| Concern Description Template | |
|---|---|
| ID | Con04 |
| Name | TollGate |
| Objective | Passing fees payment |
| Type | Functional |
| Successful scenario | See decomposition template for the TollGate concern. |
| Alternative scenario | |
| Revision date | 1st May, 2020 |
| Review count | Two times. Reviewed for tangled requirements and missing information. Therefore, the concern had to be decomposed for containing tangled descriptions of requirements. |
| Reference | Figure 2 and 3 |
| Note | This concern has several sub-concerns, which are described in a separate template. To avoid inconsistencies in describing the concern, the decomposition template provides all the necessary information related to the main and sub-concerns. |

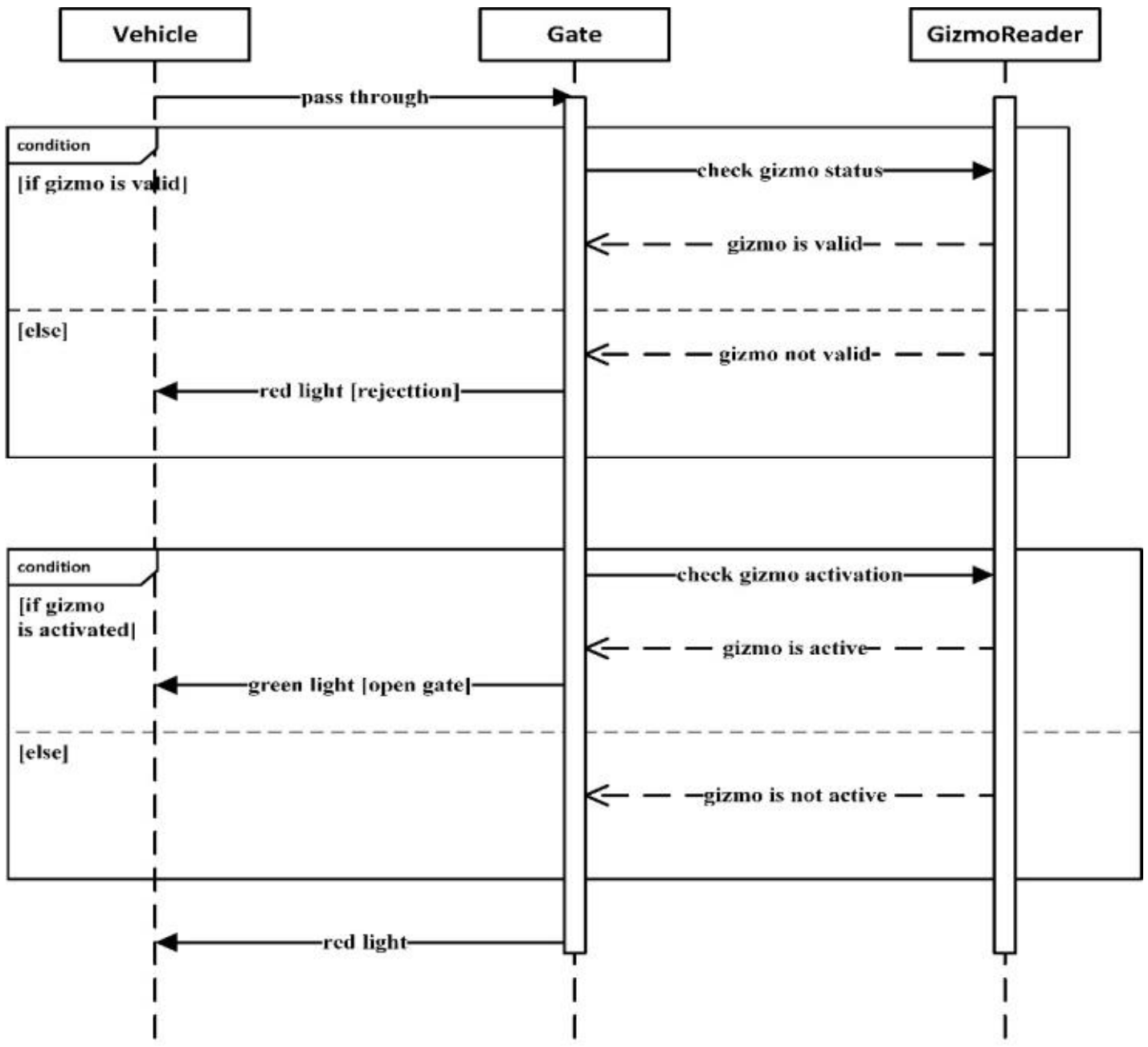

Figure 2. Demonstrates the semiformal validation of Gizmo concern

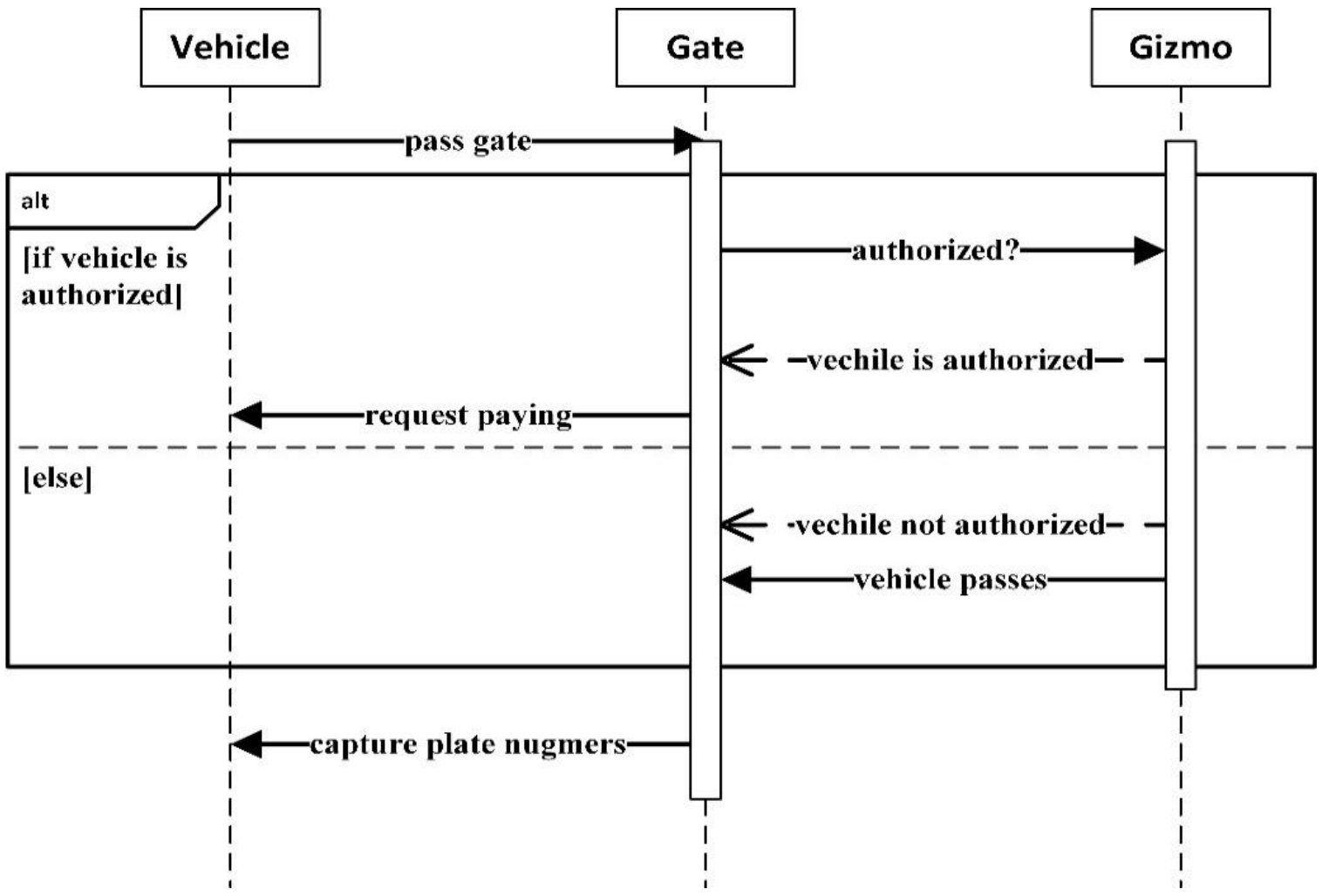

Figure 3. Demonstrates the semi-formal validation of Vehicle concern

In contrast to functional concerns, non-functional concerns cannot be validated or measured using the same techniques. Thus, we propose to document the non-functional concern in a separate template with a reference to the related functional concerns. The elements of Table 3 are explained in Activity 2 of Phase 2. The list of figures in the Reference row represents the analysis process and the influenced artefacts by *ResponseTime*. The concerns are then to be evaluated for consistency and correctness. To cope with this, we propose to use the structure in Table 4, which is a list of questions to test the consistency and correctness. This activity is explained in Activity 3 of Phase 2. This concern has been subject to several iterations of modifications. Thus, for simplicity, only the final result is presented.

Table 3. Description of ResponseTime non-functional concern

| Non-functional concern description | |
|---|---|
| Name | Response time |
| ID | Asp-01 |
| Related concerns | ATM, Gizmo, Vehicle and TollGate |
| Specification | The aspect shall respond in time to the specified concerns that are restricted by this aspect. |
| Revision date | May 4, 2020 |
| Reference | Figure 2 and 3 |

Table 4. Response time evaluation concern template

| Concern completeness | |
|---|---|
| Is the concern defined clearly? | Yes |
| Are there missing details or anything forgotten? | No |
| Is there any association with other concerns or requirements? | Yes |
| Does it include all the necessary information to make an architectural decision and design it? | Yes |
| Is there enough information to develop a test case for the concern? | Yes |
| Concern consistency | |
| Is the successful scenario of each concern written in clear language? | Yes |
| Is the description written so that it won't lead to misinterpretation? | Yes |
| Are there any parts of the concern repeated in other concerns? | Yes |
| Are there any conflicts with other concerns or requirements? | Yes |
| Is the concern traceable to its origins and stakeholder? | Yes |

Table 5. Description of ResponseTime aspectual requirement

| Aspectual Requirement Document | |
|---|---|
| ID | Asp01 |
| Name | ResponseTime |
| Concerns | ATM, Gizmo, Vehicle and TollGate |
| Description | Response time has been described and analyzed as a non-functional concern. The ATM transaction must be processed and validated immediately as the customer request the ticket from the machine in the station. Similarly, when the vehicle passes the toll gate, the system must respond quickly to the gizmo installed in the vehicle, which identifies the vehicle in terms of valid entrance, billing or authorized vehicle. |
| Priority | High |
| Pre-condition | The ATM, Gizmo, Vehicle and TollGate concern conditions must be satisfied and operate without any errors; hence the aspect will operate correctly and efficiently. |
| Reference | All functional concerns analyzed and validated in a semi-formal fashion are influenced by this aspect. |

At this stage, the aspects of the system should be identified and specified using the techniques provided by AORE approach. Moreover, the aspects are identified in different representations. As a result, we propose to document the aspectual requirements in a structured template as shown in Table 5. In most cases, there is a dependency among aspects, e.g., aspect **A** depends on aspect **B** to perform a certain task or goal. Thus, this relationship should be captured and documented so requirements engineers and analysts would have all the critical parts of each relationship. In our experiment, we did not come across any dependency between the identified aspects. The aspect then is validated to ensure that nothing important is left out of the process. For this issue, we propose to use the validation template presented in Table 6. The aspectual requirement status has to be checked against the changes made to the related concerns. As those crosscutting concerns form the aspect.

Table 6. Validation form aspectual requirements ResponseTime

| **Aspectual requirement validation** | **Aspect ID: Asp01** | **Review count: 2** |
|---|---|---|
| Does the aspect describe the solution concept for the problem? | | Yes |
| Does the aspect define the process in accurate description and understandable language? | | Yes |
| Does the aspect contain all the related parts of the crosscutting concerns? | | Yes |
| Have the areas of intersection with other requirements or concerns have been identified and specified within the scope of the aspect? | | Yes |
| Are the requirements and concerns affected by the aspect identified and represented clearly? | | Yes |

## 4.2. Theme/Doc Approach

The case study presented in the Theme/Doc approach is a course management system [6]. In this case study, a student can register and unregister for a course, the action performed by the student must be logged into the student record. The professor can register and unregister a student for a course and it must be logged in. The professor as well can give marks to the student. The lexical analysis of the requirements has addressed the following action verbs: Give, Register, Unregister and Flagged, which are functional requirements.

The functional concerns, i.e., base-themes, are modeled using a sequence diagram with the involvement of the concerned stakeholders. Figures 3 and 4 demonstrate the process of each concern that the Theme approach has produced. Figure 5 shows the clipped-view action in the Theme/Doc tool, which identifies Logged as an aspectual requirement, i.e., crosscutting the system functionalities.

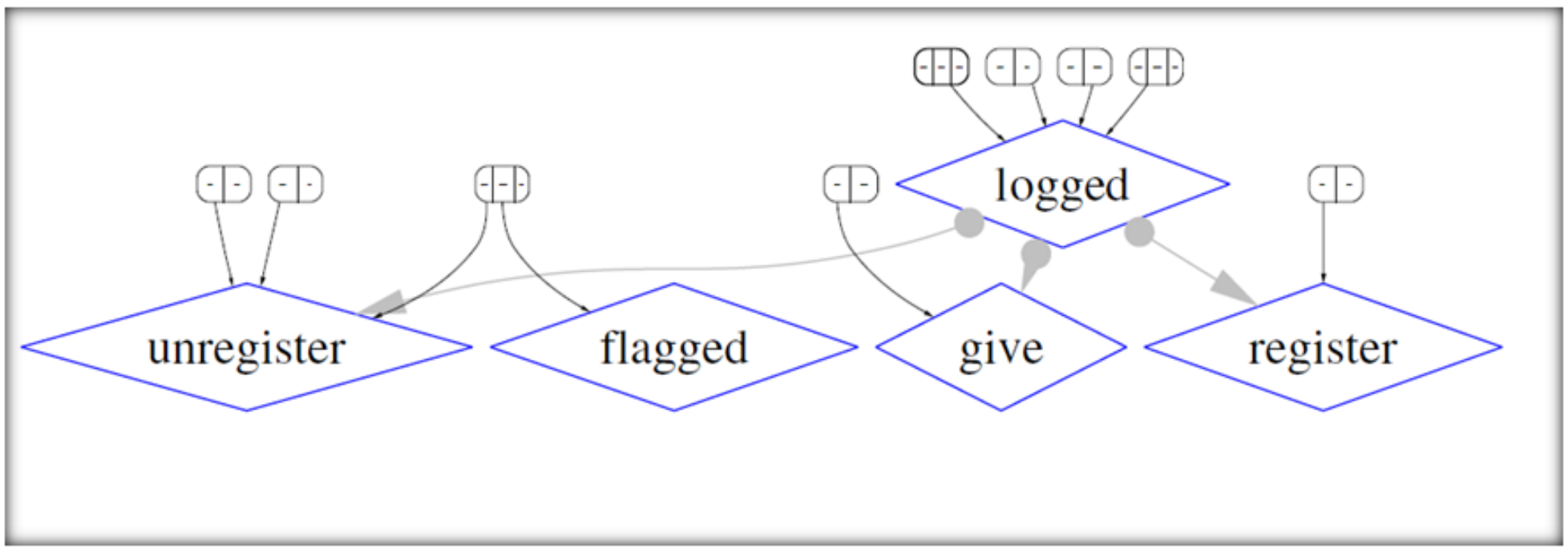

Figure 4. clipped action view of the main actions in requirements

The non-functional requirements produced by Theme/doc process in behavior, which logs each action performed by either student or professor into their records or system's database respectively. Table 9 describes the non-functional behavior, which constrains the behavior of the other functional concerns. In Table 10, the logged is evaluated against the criteria questions. There have been few refinements to the way requirements were written to formalize the last version of the evaluation template

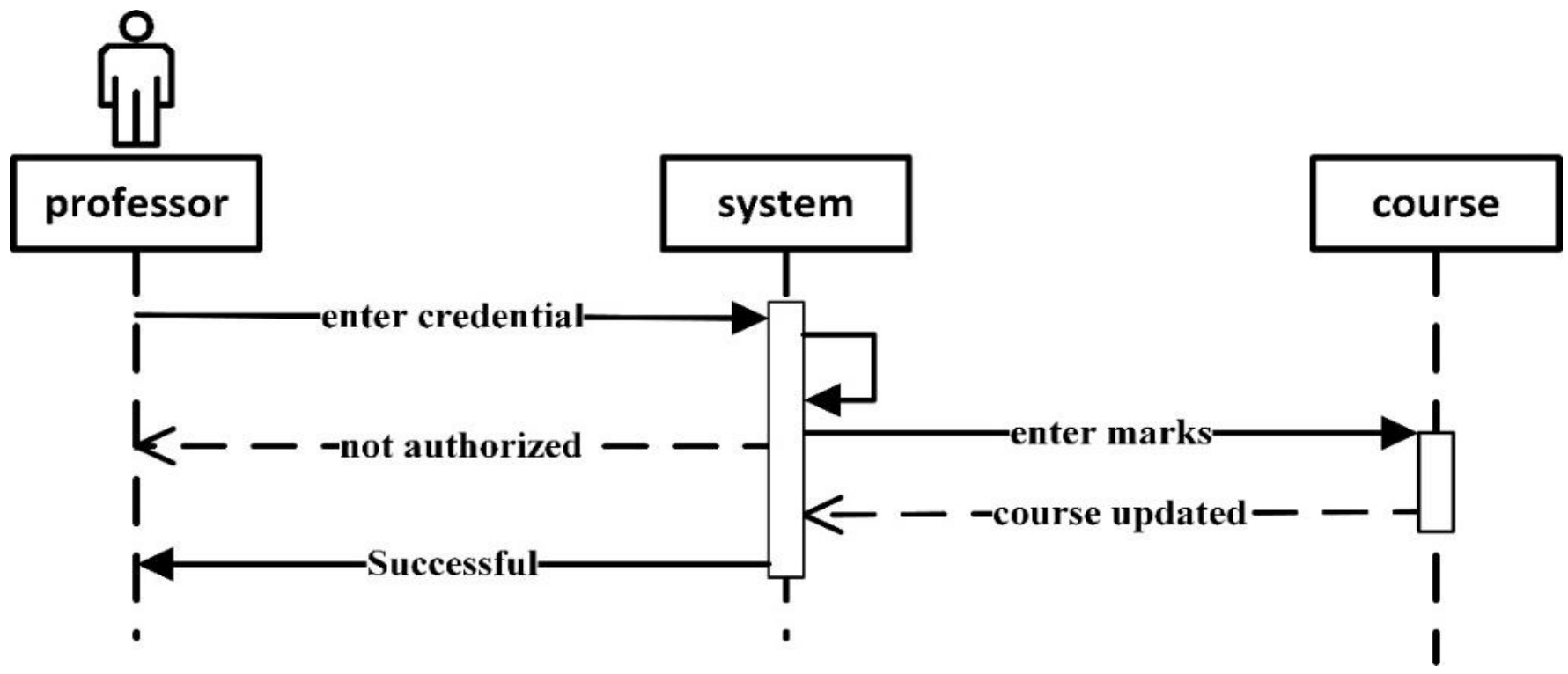

Figure 5. Sequence diagram for Give concern

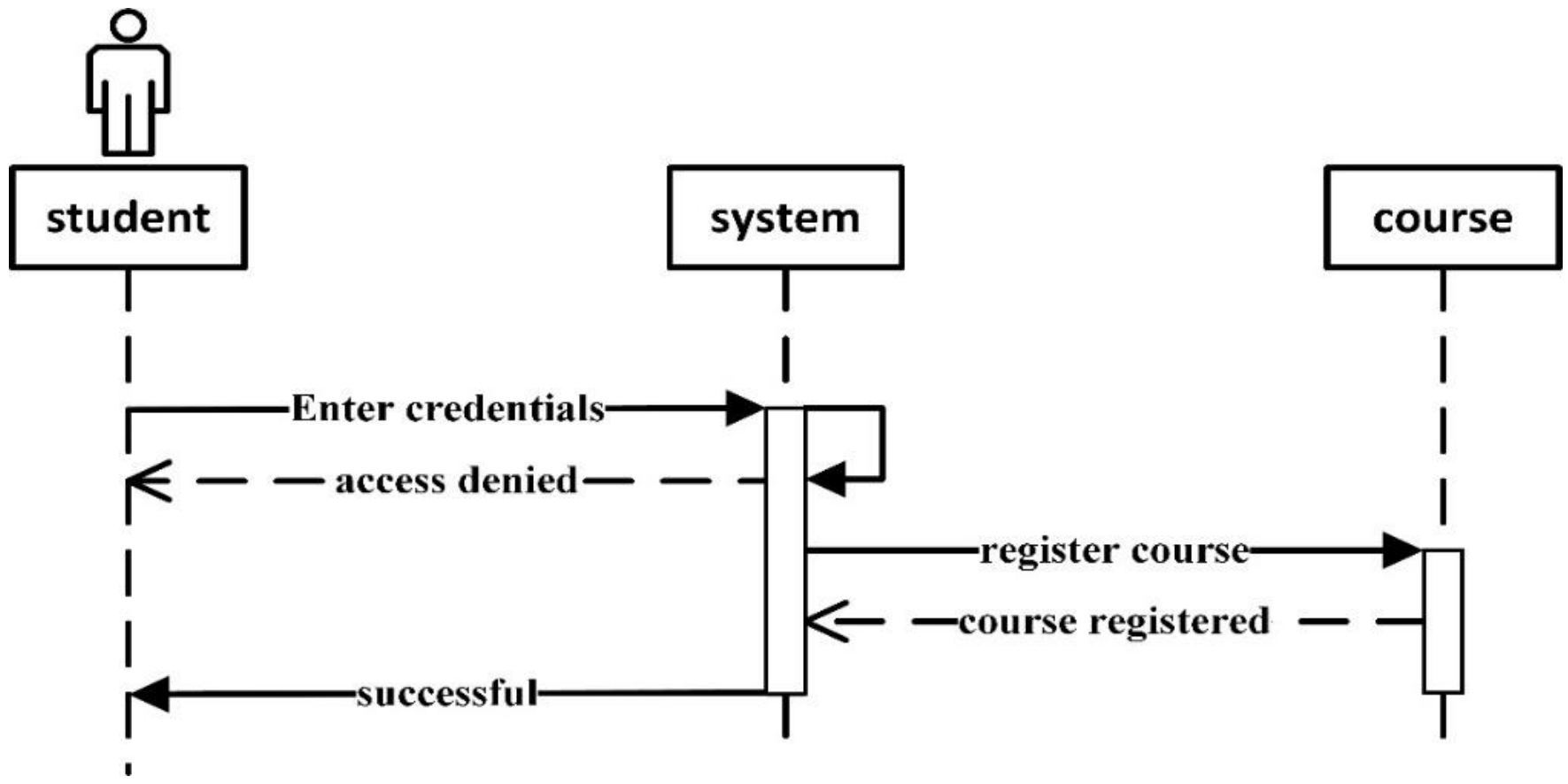

Figure 6. Sequence diagram for Register concern

The Logged requirements is a non-functional requirement. However, based on the lexical analysis of Themes, the granularity of concerns does not require the need to decompose the concerns. The description of *Give* and *Register* concerns is presented in Tables 7 and 8 respectively. The lexical nature of the analysis process has produced a solid description of requirements and concerns without the need to refine.

Table 7. Description of Give concern: base-theme

| **Concern Description Template** | |
|---|---|
| **ID** | Theme_01 |
| **Name** | Give |
| **Objective** | Enables the professor to give marks |
| **Type** | Functional |
| **Successful scenario** | The professor logs into the system and enters the marks for the specified course. |
| **Alternative scenario** | |
| **Revision date** | May 5, 2020 |
| **Review count** | 0 |
| **Reference** | Listed in system requirements in Figure. 5 and Figure. 6 |
| **Note** | This behaviour needs to be recorded into the log file. Each time the professor makes a change in the system and the database, this must be recorded. |

Table 8. Description of Register concern: base-theme

| **Concern Description Template** | |
|---|---|
| **ID** | Theme_02 |
| **Name** | Register |
| **Objective** | Enables student to register for a course |
| **Type** | |
| **Successful scenario** | The student will be asked to enter their credentials, and when credentials are correct, the system shall let students enter the course they wish to register for and submit the request. |
| **Alternative scenario** | |
| **Revision date** | May 5, 2020 |
| **Review count** | 0 |
| **Reference** | Figure 6 |
| **Note** | A log file is needed to record this activity in the system. |

Since *logged* is a non-functional concern and the non-functional behaviors are more candidates to be aspects. However, the *logged* behavior crosscuts several requirements of the system. Thus, it is an aspectual requirement. Table 11 describes the aspectual requirements based on the granularity of the information obtained in the case study conducted by the Theme/doc. Table 12 validates the aspectual requirement using the proposed checklist. The original *logged* has been reviewed two times in terms of concern relationship. Thus, as an aspectual requirement, it has to be checked against the changes made to the concern refinement.

Table 9. Description of non-functional concern *Logged*

| **Non-functional concern description** | |
|---|---|
| **Name** | Logged |
| **ID** | Theme_a |
| **Related concerns** | This concern is a potential aspect and it cuts across all the functional concerns, i.e., give marks, register student, unregister student, register course and unregister courses |
| **Specification** | Logged keeps track of the activities performed by the authorized actor. Each actor has a scope of interaction with the system. These interactions have to be logged into the individual's record. |
| **Revision date** | May 5, 2020 |
| **Reference** | Figure 4. Figure 5. Figure 6 |

Table 10. Logged concern evaluation template

| **Concern completeness** | |
|---|---|
| Is the concern defined clearly? | Yes |
| Are there missing details or anything forgotten? | No |
| Is there any association with other concerns or requirements? | Yes |
| Does it include all the necessary information to make an architectural decision and design it? | Yes |
| Is there enough information to develop a test case for the concern? | Yes |
| **Concern consistency** | |
| Is the successful scenario of each concern written in clear language? | Yes |
| Is the description written so that it won't lead to misinterpretation? | Yes |
| Are there any parts of the concern repeated in other concerns? | Yes |
| Are there any conflicts with other concerns or requirements? | Yes |
| Is the concern traceable to its origins and stakeholder? | Yes |

Table 11. Logged aspectual requirement description template

| **Aspectual Requirement Document** | |
|---|---|
| **ID** | Theme_a |
| **Name** | Logged |
| **Concerns** | Crosscuts all system concerns |
| **Description** | This aspect shall record each functionality of the system. |
| **Priority** | High |
| **Pre-condition** | The influenced concerns by this aspect must be executed and valid in terms of process. |
| **Reference** | List of requirements in requirements documents, Table 7, Table 8 and Table 9 |

Table 12. Logged aspectual requirements validation template

| **Aspectual requirement validation** | **Aspect ID: theme_a** | **Review count: 2** |
|---|---|---|
| Does the aspect describe the solution concept for the problem? | | Yes |
| Does the aspect define the process in accurate description and understandable language? | | Yes |
| Does the aspect contain all the related parts of the crosscutting concerns? | | Yes |
| Have the areas of intersection with other requirements or concerns have been identified and specified within the scope of the aspect? | | Yes |
| Are the requirements and concerns affected by the aspect identified and represented clearly? | | Yes |

## 5. DISCUSSION AND RESULTS

In this article, we have conducted two experiments on AORE approaches, i.e., AORE with ArCaDe and Theme/Doc. Theme/doc [5] represents functional concerns more effectively since they are associated with action verbs. In addition, the Theme view is a powerful tool to capture the relationship between concerns using lexical analysis of the requirements as for the non-functional concerns; the requirement documented may require being re-written. The non-functional concerns are challenging in terms of capturing the representation using Theme/doc. Using ValFAR on the concerns resulted from Theme/doc, the non-functional artefacts have gone through several iterations of refinement to avoid inconsistency of combing them with functional concerns. However, from a functional perspective, no iterations of refinement are needed to improve their granularity. Furthermore, the identified aspectual requirements in Theme/doc are not specified clearly in terms of SRS structure. Thus, we have to improve some of the themes for validation.

The Theme/Doc tool relies on lexical analysis and defining the action words to capture the relationship between base-themes, i.e., concerns, and crosscutting-themes. Moreover, the requirements must be written and documented before analysts can use the Theme/doc tool to identify the base-themes and action words that are shared with requirements to identify the crosscutting-themes, i.e., potential aspects.

However, we have found that concerns and requirements extracted using this method are more coherent and do not require any decomposition mechanism. In addition, there have been few inconsistencies describing the process of the shared requirement between the base-theme and the crosscutting-theme. This was handled by modeling the base-theme and relating it to the crosscutting-theme, which then is passed through validation for consistency and completeness to ensure the validity of the crosscutting relationship.

AORE with ArCaDe [6] uses the XML schema to store the aspect and its requirements. It represents an aspect as a concern encapsulated in XML schema, which contains the related requirements with their sub-requirements and the composition rules for the aspectual requirements. This method is effective when modularizing of aspectual requirements is highly demanded, but the treatment of concern is not efficient in terms of clarity and decomposition. By applying the ValFAR method, we discovered that requirements in the XML schema needed to be decomposed in order to validate the concerns and the aspectual requirements. This decomposition helped to achieve better granularity for each artefact. As a result, the new decomposed sub-concerns had to be analyzed and validated to maintain the cohesive structure of the schema. The decomposition rules we provided, showed that there are tangled descriptions of requirements in at least two concerns, which are identified and documented using the viewpoint approach.

## 6. CONCLUSIONS

This paper introduced a Validation Framework for Aspectual Requirements (ValFAR) and included the critical artefacts of the framework. The ValFAR method validates AORE artefacts in three phases. Each phase has its own set of activities to be performed to achieve the best results. The ValFAR enables requirements engineers to perform validation on two levels, i.e., high-level and low-level validation. The high-level validation is to validate the concerns with stakeholders and the low-level validation validates the aspectual requirement by developers using a checklist. The framework has been used to conduct an experimental study to ensure the effectiveness of the method. The experiment was conducted on the resulted artefacts of the requirements engineering phase of each approach. The results have been discussed briefly and the artefacts of ValFAR have been explained in detail. The proposed solution has shown that the results obtained from the experiments are encouraging. Furthermore, the investigation into the problem of aspect validation has shown that aspectual requirements are not specified in structured format as the regular requirements. As a result, developers document the aspects to the best of their knowledge without fully grasping the concepts of crosscutting concerns and aspects. To conclude, the approaches discussed in the related work showed that the existing approaches specify concerns and aspects to their satisfaction. Each approach describes and specifies aspects differently from other approaches. Thus, a standard and formal specification of aspects are desired to cope with this issue.

The proposed solution treats each viewpoint and theme as a concern, which facilitates the process of concern decomposition to validate the crosscutting relationship by inspecting each concern individually. This benefits in determining the correctness and consistency of crosscutting relationship components. The resulted aspectual requirements are validated with a checklist, which aims at defining what aspect should be in the requirements phase. The evaluation of the proposed solution on the different approaches has shown that the resulting aspectual requirement requires more refinement and a deep understanding of the nature of each one to validate them.
We have demonstrated the ValFAR method with two different AORE approaches step by step. This presents a limitation to the work we conducted. Furthermore, we believe that our work might be carried out with other approaches, which produce different AORE artefacts. The main idea is to handle each artefact as a crosscutting concern, i.e., aspectual requirements. Thus, we encourage conducting more case studies using ValFAR.

Finally, we are working on validating the integrity and the correctness of the proposed solution with formal methods. Formalizing the framework would ensure its correctness and accuracy as well as testing it for shortcomings. On the other hand, the demonstration of the work in this paper is presented manually, and therefore, requirements engineers and analysts have to pay attention to every little detail along the process. Moreover, future work is concentrating on automating this framework in a software tool to increase time and effort productivity for ease of use.